\documentclass[twocolumn,pra,aps,superscriptaddress,showpacs,10pt]{revtex4-2}

\usepackage{amsmath}
\usepackage{amssymb}
\usepackage{graphicx}
\usepackage{subfig}
\graphicspath{{fig/}}

\usepackage{color}

\newcommand{\ket}[1]{\left|#1\right\rangle}
\newcommand{\figref}[1]{FIG.~\ref{#1}}

\begin{document}

\title{A Framework for Quantum Ray Tracing}
  
\author{Xi Lu}
\affiliation{School of Mathematical Science, Zhejiang University, Hangzhou, 310027, China}
\affiliation{State Key Lab. of CAD\&CG, Zhejiang University, Hangzhou, 310058, China}

\author{Hongwei Lin}
\email{hwlin@zju.edu.cn}
\affiliation{School of Mathematical Science, Zhejiang University, Hangzhou, 310027, China}
\affiliation{State Key Lab. of CAD\&CG, Zhejiang University, Hangzhou, 310058, China}

\begin{abstract}
  Ray tracing algorithm simulates the physical movements of a huge amount of rays to render a high quality image, in which the tracing procedure for each ray can be implemented in parallel.
  By leveraging the inherent parallelism of quantum computing, we propose a quantum ray tracing algorithm, which is proved to have a quadratic speedup over the classical path tracing.
\end{abstract}

\maketitle


\section{Introduction}
\label{sec:introduction}

The ray tracing algorithm\cite{Whitted1980, Cook1984, Kajiya1986, haines2019ray} is a general term of rendering algorithms that calculate pixel colors by simulating the physical interactions of light rays and the scene, such as reflections and refractions.
To render an image with high quality, ray tracing algorithms require simulating an astronomical number of rays.
To be specific, a single ray scatters towards many directions when interacting with an object, and each of the scattered rays scatters towards more directions when interacting with other objects, so the total number of rays grow exponentially in the number of interactions.
In many situations people have to make a trade-off between time cost and quality.
For real-time ray tracing applications where the rendering duration is strictly limited, the state-of-the-art GPU can only handle sampling a small amount of rays per pixel, and the resultant noise is fixed by a subsequent denoising procedure\cite{marrs2021ray}.

In this paper we focus on the ray tracing algorithm on quantum computers.
Quantum computation is an emerging subject that studies how to perform computational tasks in quantum mechanical systems.
By leveraging the superposition and entanglement of quantum computing, quantum computing has inherent advantages on parallel computational tasks.
As a result, quantum computing shows it computational power by providing spectacular speedup over classical computing in some problems\cite{Grover1997, Shor1997}.

The ray tracing algorithm can be easily parallelized, since the procedure for tracing each ray is the same.
Then comes an interesting question: is the inherently parallel quantum computing able to speed up the inherently parallel ray tracing algorithm?
This paper will give a positive answer as well as a fully practicable implementation.


The idea of introducing quantum computing into rendering methods such as Z-buffer, ray tracing and radiosity algorithm was proposed by \cite{Lanzagorta2005}, which presents the concepts but does not give a fully practicable implementation.
In particular, they proposed a quantum ray tracing solution by superposing all scene primitives.
The idea of using the quantum parallelism property in computer graphics was practiced in \cite{Eric2016, Shimada2020}, which both used an amplitude amplification\cite{Abrams1999} based quantum sum estimation and applied their methods on filtering binary images.
The paper \cite{Alves2019} proposed an implementation of Grover's algorithm for ray casting from an orthographic camera.

\begin{figure*}
    \centering
    \subfloat[Classical path tracing.]{
        \includegraphics[width=.4\linewidth]{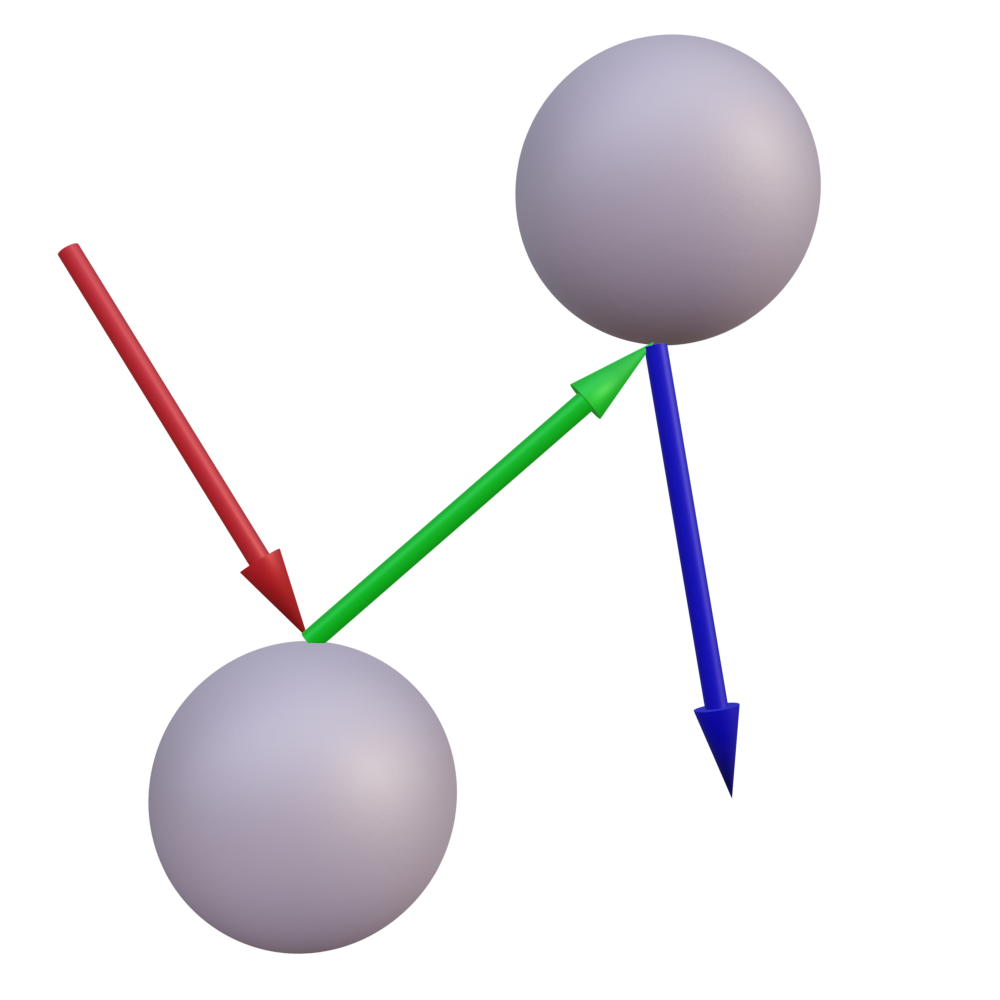}
        \label{fig:classical-ray-tracing}
    }
    \qquad
    \subfloat[Quantum ray tracing.]{
        \includegraphics[width=.4\linewidth]{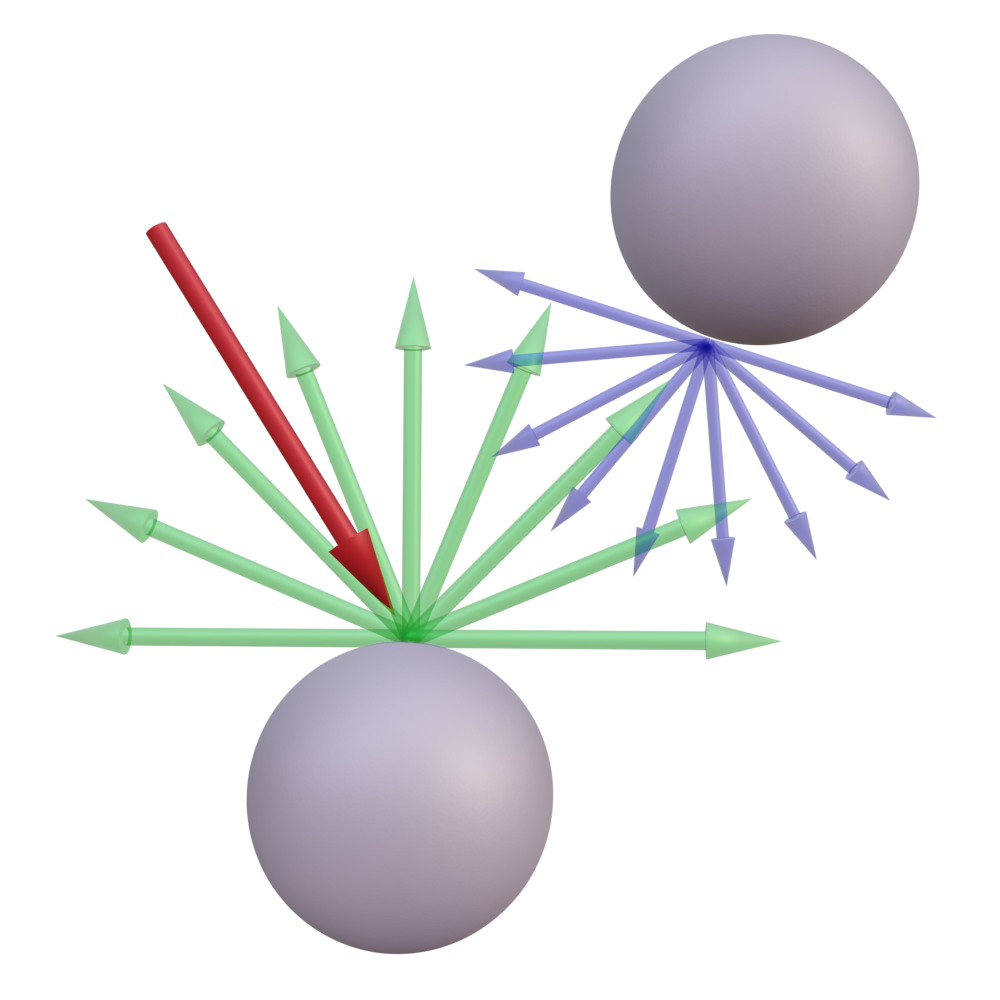}
        \label{fig:quantum-ray-tracing}
    }
    \caption{Classical path tracing only traces one ray at a time, while quantum ray tracing can trace numerous rays as a superposition in one shot.}
    \label{fig:ray-tracing}
\end{figure*}

The structure of this paper is as follows.
Section~\ref{sec:preliminary} briefly introduces both quantum computing and classical ray tracing.
In section~\ref{sec:algorithm}, we first give a big picture for our quantum ray tracing algorithm, then dive into three details including how to perform dense sampling, how rays interact with the scene and how to extract the average color from a superposition as final output.
In section~\ref{sec:performance}, we theoretically analyze the space and time complexity of quantum ray tracing, then make a comparison between quantum ray tracing and classical ray tracing, in the sense of achieving the same deviation.
Finally, in section \ref{sec:discussion}, we make some conclusions and discussions.


\section{Preliminary}
\label{sec:preliminary}

\subsection{Quantum Computing}

All stories began in the 1980s when Feynman suggested that quantum mechanics might be more computationally powerful than Turing mechine\cite{Feynman1982,Feynman1986}.
By substituting classical bits for quantum bits, or {\itshape qubits}, which can be not only in the states $\ket{0}$ and $\ket{1}$ but also their superposition $a\ket{0}+b\ket{1}$ where $a,b\in\mathbb{C}$ and $|a|^2+|b|^2=1$, quantum computing obtains many interesting features like entanglement, reversibility, parallelism, no-cloning and indistinguishability.

In quantum computing, operations on qubits are implemented by {\itshape quantum gates}.
There are two classes of gates, namely unitary gates and measurement gates.
Unitary gates perform unitary transformations to the state vectors, while measurement gates perform probabilistic and destructive transformations to extract information from quantum states.
A {\itshape quantum circuit} is said to be a group of quantum gates that can perform specific functions.

A quantum computer can simulate a classical computer, by restricting the qubit states to $\{\ket{0},\ket{1}\}$, and using the Toffoli gate, X gate and CNOT gate to replace the AND gate, NOT gate and the copy operation in classical computers, respectively.
\begin{equation*}
    \begin{aligned}
        \text{Toffoli: }& \ket{a}\ket{b}\ket{0} &\mapsto& \ket{a}\ket{b}\ket{a\text{ and }b} \\
        \text{X: }& \ket{a} &\mapsto& \ket{\text{not }a} \\
        \text{CNOT: }& \ket{a}\ket{0} &\mapsto& \ket{a}\ket{a}
    \end{aligned}
\end{equation*}

Furthermore, due to the reversibility of the three gates above, the quantum implementation of a classical function $j\mapsto f(j)$ should be of the following form,
\begin{equation}
    \ket{j}\ket{0} \mapsto \ket{j}\ket{f(j)},
\end{equation}
which is sometimes abbreviated as $\ket{j}\mapsto\ket{j}\ket{f(j)}$.
Here $\ket{j}$ and $\ket{f(j)}$ are quantum registers that use several qubits to store various data structures like integers and real numbers.

If we avoid using any measurement gate in the circuit of computing $f$, then we can make full use of the linearity and reversibility of unitary gates.
It follows immediately that when a superposition state is inputted, the same circuit performs the following linear transformation,
\begin{equation}
    \sum_jx_j\ket{j} \mapsto \sum_jx_j\ket{j}\ket{f(j)},
    \label{eq:oracle-f}
\end{equation}
due to the linear property.
We call such circuits {\itshape linear circuits}.

It seems that several evaluations of the function $f$ can be obtained in one query.
But once we have access to a specific $f(j_0)$, no matter by which means, the whole state must collapse to the basis state $\ket{j_0}\ket{f(j_0)}$, and the information of other evaluations is lost forever.
Anyway, we have to design clever algorithms to make the best use of quantum parallelism.
Some of such examples are Grover's search\cite{Grover1997}, minimum finding\cite{Durr1996}, quantum counting\cite{QuantumCounting1998}, and quantum numerical integrals\cite{Abrams1999}.


\subsection{Classical Ray Tracing}

To calculate the color, or the ray energy, emitted by light sources and received by the camera, the ray tracing algorithm utilizes the reversibility of light ray paths, that is, shoots rays from the camera, simulates the physical interactions between rays and scene objects, until they hit the light sources.
The core mathematical problem in ray tracing is to solve the {\itshape rendering equation}\cite{Kajiya1986} for computing the light radiance from an object surface in a certain direction,
\begin{equation}
    L_o(\mathbf{r}_o) = L_e(\mathbf{r}_o) + \int_\Omega L_i(\mathbf{r}_i)f_\text{BSDF}(\mathbf{r}_i,\mathbf{r}_o)(\mathbf{n}\cdot\mathbf{r}_i)\operatorname{d}\mathbf{r}_i,
    \label{eq:rendering-equation}
\end{equation}
where the integral domain $\Omega$ is the unit sphere, $L_o, L_i, L_e$ stands for the outgoing, incoming, self-emission radiance respectively, and $f_\text{BSDF}$ is the {\itshape bidirectional scattering distribution function} that is related to the material of the object.
This spherical integral takes into account the contributions of the reflected or refracted rays in all directions.

Observe that the $L_i$ term in the integrand is equal to some $L_o$ in another rendering equation, thus Eq.~\eqref{eq:rendering-equation} is infinitely recursive.
A common solution is to use a Russian Roulette at each depth to decide whether to terminate, and make corresponding compensation.
Another rough solution is to pre-set an upper bound $D$, and the recursion is halted when reaching depth $D$.

The standard solution in classical ray tracing to solve the rendering equation is Monte Carlo sampling.
One approach is the {\itshape path tracing}\cite{Kajiya1986}, in which only one randomly chosen ray is shot outwards whenever a ray hits an object, as illustrated in \figref{fig:ray-tracing}(a).
For each pixel many paths are traced, and the final color written to that pixel is the average color of these paths.


\section{Quantum Ray Tracing}
\label{sec:algorithm}

\begin{figure*}
    \centering
    \includegraphics[width=\linewidth]{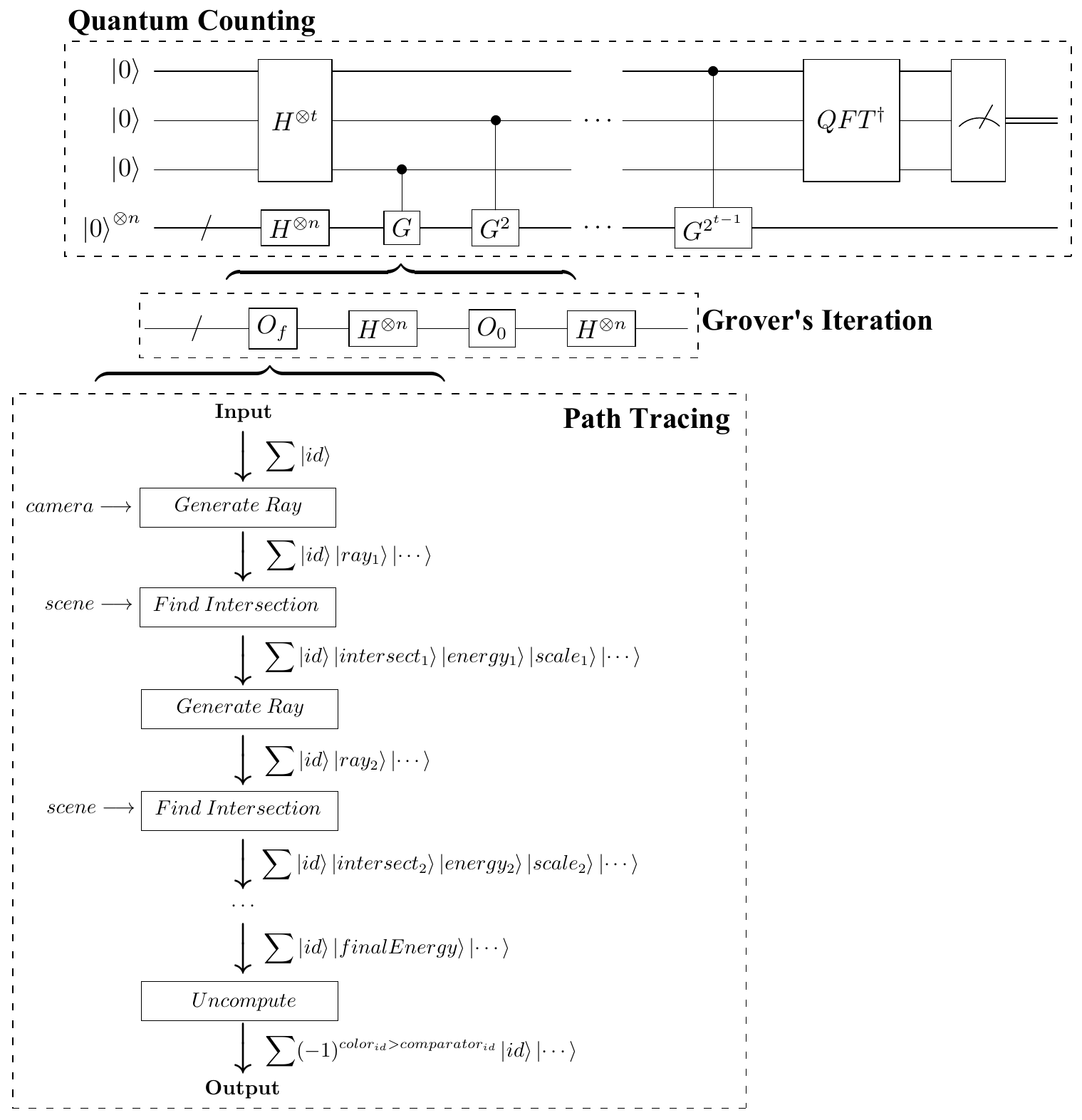}
    \caption{A macrostructure for our quantum ray tracing algorithm. The \textbf{Quantum Counting} frame illustrates the circuit of the well-known quantum counting algorithm. The $G$ gate performs a Grover's iteration, which is shown in the \textbf{Grover's Iteration} frame. The $O_0$ gate shifts the phase of every computational basis except $\ket{00\cdots0}$. The $O_f$ gate performs the whole path tracing algorithm, whose details are shown in the \textbf{Path Tracing} frame. }
    \label{fig:big-picture}
\end{figure*}

For a single pixel on the camera, classical ray tracing shoots many rays and calculates their average energy as the output color.
The key idea of quantum ray tracing is to store all those rays in a superposition.
In quantum computing, if we store the rays in the following form,
\begin{equation}
    \sum_{id=0}^{N-1} x_{id} \ket{id} \ket{ray_{id}},
    \label{eq:ray-form}
\end{equation}
where $N$ is the total number of superposed rays, $x_{id}\in\mathbb{C}$ are complex coefficients, and $\ket{ray_{id}}$ is a structured register that stores the origin and the direction information of the ray, then by the idea of quantum linear circuit we can trace all those rays in one shot.

In classical ray tracing, if we shoot $R_k$ rays outwards at the $k$-th interaction, then a total number of $O(R)$ space and time are required, where $R=\prod_{k=1}^DR_k$ if we trace rays to depth $D$.
That is why classical path tracing sets $R_k=1$ to avoid exponential explosion.
But in quantum computing, according to Eq.~\ref{eq:ray-form}, only $O(\log R)$ space that stores $\ket{id}$ are required, and the subsequent procedure for tracing the superposed state of rays with a quantum linear circuit is the same as tracing a single ray.
The details of how to implement ray-object interactions are discussed in Section~\ref{sec:interaction}.

We can feel free to choose a large $R$ in quantum computing, that is, sample a densely distributed directions at each interaction.
At each interaction we append a new quantum register to store the newly shot ray, and after tracing to depth $D$ the state becomes,
\begin{equation}
    \begin{aligned}
        \sum_{id} & \ket{id} \ket{primaryRay_{id}} \ket{secondaryRay_{id}} \cdots \\ & \ket{lastRay_{id}} \ket{\cdots},
    \end{aligned}
\end{equation}
where the coefficients are abbreviated for the convenience of writing, and $\ket{\cdots}$ stands for possible garbage registers.
The details of obtaining such state are discussed in Section~\ref{sec:sampling}.

Finally, each $id$ stands for a single ray path to depth $D$.
During the tracing procedure we use auxiliary registers to store the accumulated energy and the energy scale at each interaction.
In the end there is a register that stores the total ray energy to depth $D$, we denote the final state as,
\begin{equation}
    \sum_{id} \ket{id} \ket{color_{id}} \ket{\cdots}.
    \label{eq:color-to-average}
\end{equation}

Having obtained the final energy of each path, the only thing left is to calculate their average, to get the final color of the pixel.
Unfortunately, the energy information is entangled, and we can never read all of them from a single state.
Hence, we need to use an algorithm to extract information from some repetitions of all procedures above.
The quantum averaging algorithm in Section~\ref{sec:averaging} is a quantum counting\cite{QuantumCounting1998} based algorithm that construct a Boolean function to deal with the real numbers in register $color$ and use the standard quantum counting algorithm to estimate their average.
Since the outcome of quantum averaging algorithm is a single real number, and the RGB model of a color contains three numbers, we should run the whole procedure for each pixel and each RGB channel to render the whole image.

A macrostructure for our quantum ray tracing algorithm is shown in \figref{fig:big-picture}.

\subsection{Dense sampling}
\label{sec:sampling}

In this part we discuss the \textbf{Generate Ray} steps in \figref{fig:big-picture}.
The sampling happens at each depth of ray-object interaction, thus we need to prepare a superposition ID for each depth.
We divide the register $id$ into several parts, and each of them works for only one sampling step,
\begin{equation}
    \begin{aligned}
        \ket{id} = & \ket{primaryRayId} \ket{secondarRayId} \cdots
        \\ & \ket{finalRayId} \ket{comparatorId},
        \label{eq:id-register}
    \end{aligned}
\end{equation}
where the utility of $\ket{comparatorId}$ will be discussed in Section~\ref{sec:averaging}.

Given a pixel square, the primary rays are constructed by setting the origin to the world position of the camera, and calculating the direction according to the world rotation of the camera and the screen position on the camera.
In classical ray tracing, the screen position of rays are uniformly randomly distributed within the corresponding pixel for anti-aliasing.
Here in quantum ray tracing, we replace the random sampling with a dense superposed sampling,
\begin{equation}
    \begin{aligned}
        & \sum\ket{primaryRayId}\ket{\cdots} 
        \\ \mapsto & \sum\ket{primaryRayId}\ket{primaryRay} \ket{\cdots}.
    \end{aligned}
\end{equation}

\begin{figure}[ht]
    \centering
    \includegraphics[width=\linewidth]{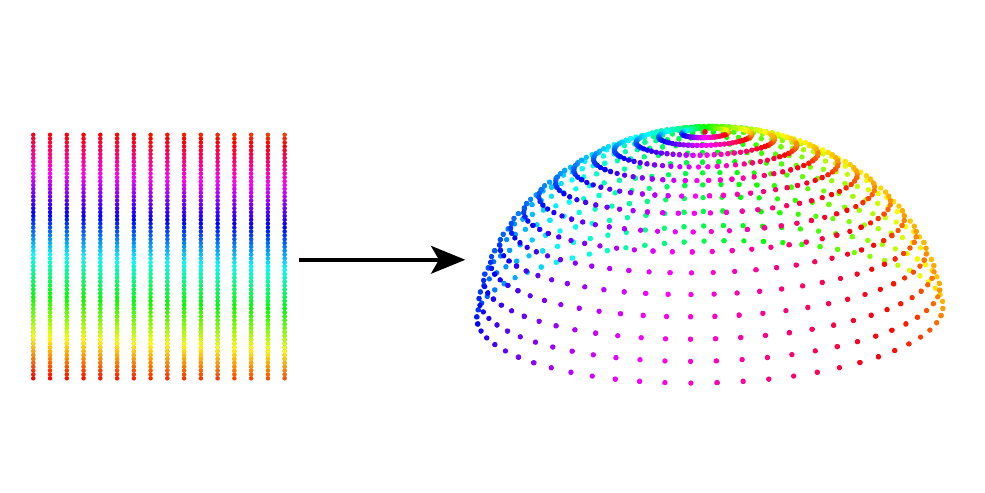}
    \caption{Sample a hemisphere $\Omega_+$ in superposition from a lattice.}
    \label{fig:sample-hemisphere}
\end{figure}

From secondary ray on, the sampling happens when rays interact with scene objects, and the sampling domains are unit spheres, or unit hemispheres when objects are opaque.
We can first sample a 2D lattice, then map it onto our desired domain, as illustrated in \figref{fig:sample-hemisphere}.

The random sampling approach for calculating the numerical integral uses the following approximation,
\begin{equation}
    \begin{aligned}
        & \int_{\Omega_+} L(\mathbf{r})f_\text{BSDF}(\mathbf{r},\mathbf{r}_o)\cdot (\mathbf{n}\cdot\mathbf{r})\operatorname{d}\mathbf{r}
        \\ \approx & \frac{1}{R_k} \sum_{j=1}^{R_k} \frac{L(\mathbf{r}_j)f_\text{BSDF}(\mathbf{r}_j,\mathbf{r}_o)\cdot (\mathbf{n}\cdot\mathbf{r}_j)}{p(\mathbf{r}_j)},
    \end{aligned}
    \label{eq:classical-integral}
\end{equation}
where $R_k$ is the number of samples, $\{\mathbf{r}_j\}$ are sampled from the integral domain $\Omega_+$ with respect to the probability density function $p$.
Here in dense sampling for quantum ray tracing, we use a smooth mapping $\phi:[0,1]^2\rightarrow\Omega_+$ to replace the random distribution $p$, hence Eq.~\eqref{eq:classical-integral} should be replaced by,
\begin{equation}
    \begin{aligned}
        & \int_{\Omega_+} L(\mathbf{r})f_\text{BSDF}(\mathbf{r},\mathbf{r}_o)\cdot (\mathbf{n}\cdot\mathbf{r})\operatorname{d}\mathbf{r}
        \\ \approx & \frac{1}{R_k} \sum_{j=1}^{R_k} \frac{L(\mathbf{r}_j)f_\text{BSDF}(\mathbf{r}_j,\mathbf{r}_o)\cdot (\mathbf{n}\cdot\mathbf{r}_j)}{\det(\operatorname{D}\phi(x_j))}.
    \end{aligned}
    \label{eq:quantum-integral}
\end{equation}
where $\{x_j\}$ here are lattice points, $\operatorname{D}\phi$ is the differential of $\phi$, and $\det(\operatorname{D}\phi)$ is its determinant.

\subsection{Interaction with scene}
\label{sec:interaction}

In this part we discuss the \textbf{Find Interaction} step in \figref{fig:big-picture}.
The whole scene is inputted as a list of primitives, for example triangles.
For each triangle, a ray-triangle intersection test is implemented, and the results are stored in the structured register $intersect$ that contains information including whether the interaction exist, and the distance, position, normal, texture coordinate and material ID of the interaction, in the following form,
\begin{equation}
    \begin{aligned}
        \sum_{id} & \ket{id} \ket{ray} \ket{\cdots}
        \\ & \ket{intersect_1} \ket{interact_2} \cdots \ket{intersect_P},
    \end{aligned}
\end{equation}
where $P$ is the number of primitives, and $intersect_k(k=1,2,\cdots,p)$ stores the intersection information of a ray and the $k$-th primitive.

From those intersections, only the one that exists and has the smallest distance should be picked out.
Define $nearest_k(k=1,2,\cdots,p)$ to be the nearest intersection in $\{intersect_j:1\le j\le k\}$, then we should build a chain that picks out the nearer intersection between $intersect_k$ and $nearest_{k-1}$ to decide which should be copied to $nearest_k$, where $k=2,3,\cdots,P$, to obtain the state,
\begin{equation}
    \begin{aligned}
        \sum_{id} & \ket{id} \ket{ray} \ket{\cdots}
        \\ & \ket{intersect_1} \ket{interact_2} \cdots \ket{intersect_P}
        \\ & \ket{nearest_2} \cdots \ket{nearest_P}.
    \end{aligned}
\end{equation}

Finally, $nearest_P$ is the desired intersection between ray and scene.
We abbreviate the current state as,
\begin{equation}
    \sum \ket{id} \ket{ray} \ket{intersect} \ket{\cdots}.
\end{equation}

To compute the accumulated ray energy of each superposed path, one problem is that different rays are interacting with different materials.
Since the intersection structure contains a member that stores material ID, we traverse all materials in scene, and whether a path and a material interact is controlled by whether their material IDs meet.
Since the number of materials is no more than the number of primitives, the total space and time complexity of performing a whole \textbf{Find Intersection} procedure are $O(P)$.

\subsection{Computing the average color}
\label{sec:averaging}

The idea of quantum sum estimation comes from the quantum counting algorithm\cite{QuantumCounting1998}.
Given a Boolean function $f: \{0, 1, \cdots, N-1\} \rightarrow \{0, 1\}$ where it is assumed that $N=2^n(n\in\mathbb{Z}_+)$ without loss of generality, and a corresponding phase oracle,
\begin{equation}
    O_f: \sum_{id}x_{id}\ket{id} \mapsto \sum_{id}(-1)^{f(id)}x_{id}\ket{id},
    \label{eq:phase-flip}
\end{equation}
the quantum counting algorithm can output an estimation $\tilde{S}$ of the sum $S=\sum_{j=0}^{N-1}f(j)$, such that 
\begin{equation}
    |\tilde{S}-S| < \frac{2\pi\sqrt{S}}{T}+\frac{\pi^2}{T^2},
\end{equation}
with probability at least $8/\pi^2$\cite{QuantumCounting1998}, where $T=2^t$, and $t$ is the number of qubits in the first register in the \textbf{Quantum Counting} frame in \figref{fig:big-picture}.

The idea of quantum counting is to estimate the eigenvalues of the unitary transformation of a Grover's iteration\cite{Grover1997} for Boolean function $f$, namely $e^{2\pi i\theta}$ and $e^{-2\pi i\theta}$, where
\begin{equation}
    \theta = \frac{1}{\pi}\arcsin\sqrt{\frac{S}{N}} \in \left[0,\frac{1}{2}\right].
    \label{eq:phase-fraction}
\end{equation}

\begin{figure}[ht]
    \centering
    \includegraphics[width=\linewidth]{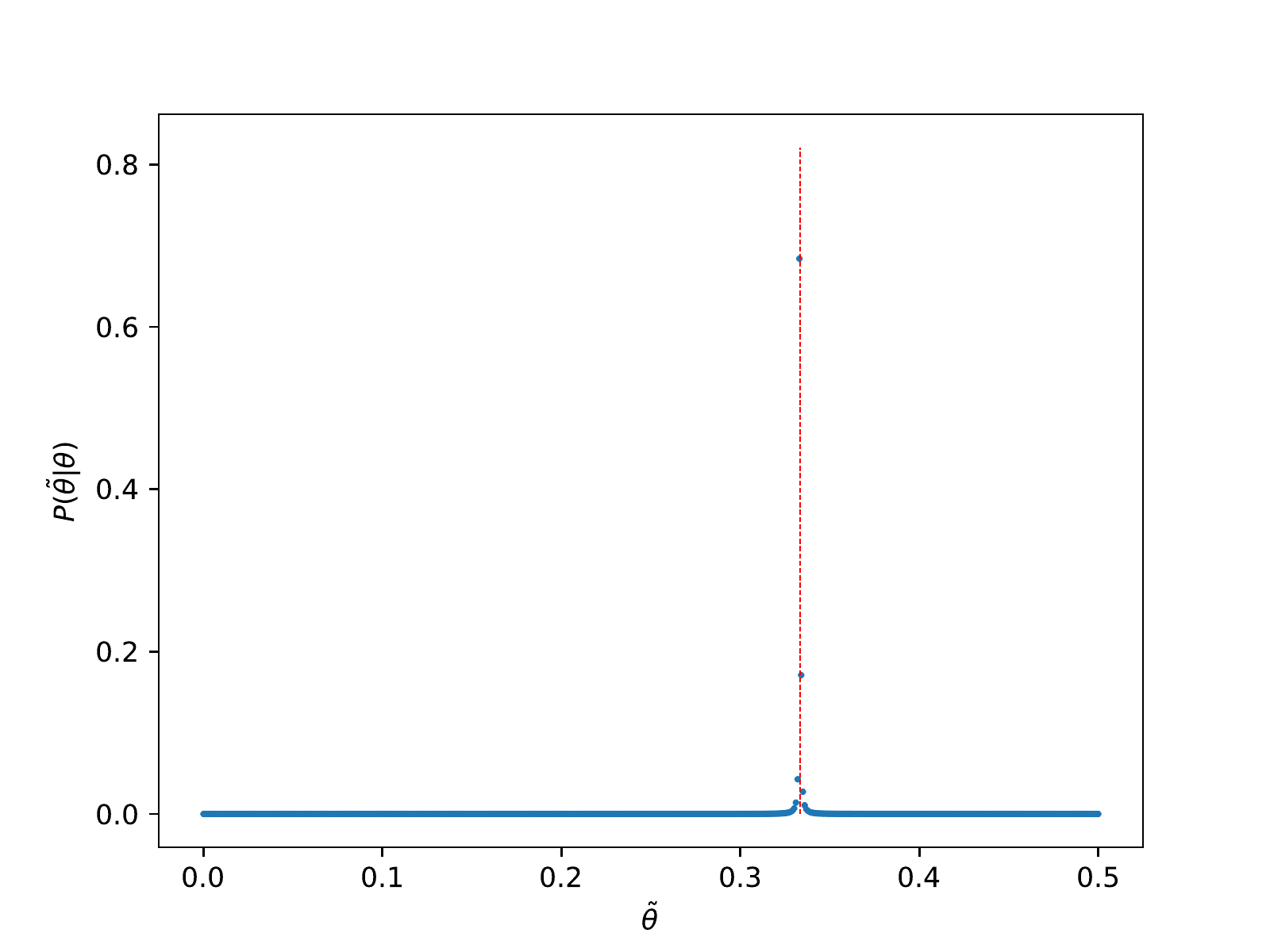}
    \caption{A graph for the probability distribution $P(\tilde{\theta}|\theta)$, in which $T=1024$ and $\theta=1/3$. The red vertical line shows the value of $\theta$.}
    \label{fig:p-theta}
\end{figure}

The quantum counting algorithm can output a discrete random variable $\tilde{\theta}$ with distribution,
\begin{equation}
    P(\tilde{\theta}|\theta) = \begin{cases}
        \left(\frac{\sin(T\pi(\tilde{\theta}-\theta))}{T\sin(\pi(\tilde{\theta}-\theta))}\right)^2 + & \left(\frac{\sin(T\pi(\tilde{\theta}+\theta))}{T\sin(\pi(\tilde{\theta}+\theta))}\right)^2, \\
        & \tilde{\theta}=\frac{1}{T},\cdots, \frac{T/2-1}{T}; \\
        \left(\frac{\sin(T\pi(\tilde{\theta}-\theta))}{T\sin(\pi(\tilde{\theta}-\theta))}\right)^2, & \tilde{\theta} = 0,\frac{1}{2};
    \end{cases}
\end{equation}
which has a sharp peak around $\tilde{\theta}=\theta$ for large $T$, as illustrated in \figref{fig:p-theta}.

Practically, we can repeat the phase estimation for $B$ times, to obtain a result set $\{\tilde{\theta}_k\}_{k=1}^{B}$.
Since $\tilde{\theta}$ is not an unbiased estimation of $\theta$, a better way than taking an average is to use Bayesian estimation,
\begin{equation}
    P(\theta| \{\tilde{\theta}_k\}) = \frac{P(\{\tilde{\theta}_k\}| \theta)P(\theta)}{\sum_{\theta'} P(\{\tilde{\theta}_k\}| \theta')P(\theta')},
\end{equation}
to estimate $\theta$ and thus $S$.
According to Eq.~\eqref{eq:phase-fraction}, the possible values of $\theta$ are discrete.
Assuming $\theta$ is evenly distributed, that is, all $P(\theta)$ are equal.
Then by finding the maximum of Eq.~\eqref{eq:Bayesian} we obtain the final estimation $\tilde{\theta}$ of $\theta$.
\begin{equation}
    P(\theta| \{\tilde{\theta}_k\}) \propto P(\{\tilde{\theta}_k\}| \theta) = \prod_{k=1}^{B}P(\tilde{\theta}_k| \theta).
    \label{eq:Bayesian}
\end{equation}

Now we have the state Eq.~\eqref{eq:color-to-average}, where each $color_{id}$ is a non-negative real number, and is entangled with a unique $id$.
Our goal is to estimate their average.

Suppose we hope to calculate the average of a set of numbers $x\in\mathcal{X}$ bounded by the range $[0, 2^b)$.
We first build a comparator set,
\begin{equation}
    \mathcal{Y} = \left\{ 2^{b-c}y | y=0,1,\cdots,2^c-1 \right\},
\end{equation}
and a comparison function,
\begin{equation}
    f(x, y) = \begin{cases}
        1, & x > y;\\
        0, & x \le y.
    \end{cases}
\end{equation}

Then the average of the summands $x\in\mathcal{X}$ can be approximated via,
\begin{equation}
    \left| \frac{1}{|\mathcal{X}|}\sum_{x\in\mathcal{X}}x - \frac{2^{b-c}}{|\mathcal{X}|}\sum_{x\in\mathcal{X},y\in\mathcal{Y}}f(x,y) \right| < 2^{b-c-1}.
\end{equation}

Additionally, if $x$ is stored in a fixed-point format with total bit length $c$ and integer bit length $b$, then the approximation above is exact.

Remember that one part of the $id$ register is spared for storing $comparatorId$.
Suppose the $comparatorId$ register consists of $c$ qubits, then each path is entangled with a superposition
\begin{equation}
    \sum_{comparatorId=0}^{C-1}\ket{comparatorId},
\end{equation}
where $C=2^c$.
Then we use the quantity,
\begin{equation}
    2^{b-n}\sum_{id}f(color_{id},comparator_{id}),
\end{equation}
to approximate the average of colors, where $n$ is the size of $id$ register.
Here $f$ is a Boolean function, hence the sum can be approximated by the standard quantum counting algorithm.
We apply phases shifts to the $id$s that satisfies $color_{id} > comparator_{id}$, that is, perform the following transformation,
\begin{equation}
    \begin{aligned}
        & \sum_{id} \ket{id} \ket{color_{id}} \ket{comparator_{id}} \ket{\cdots} 
        \\ \mapsto & \sum_{id} (-1)^{f(color_{id}, comparator_{id})} \ket{id} 
        \\ & \ket{color_{id}} \ket{comparator_{id}} \ket{\cdots}.
    \end{aligned}
\end{equation}

Finally, an uncomputing procedure, which inverses all circuit above except the final phase shifting, is performed to obtain the state,
\begin{equation}
    \sum_{id} \ket{id} \mapsto \sum_{id}(-1)^{f(color_{id}, comparator_{id})} \ket{id},
\end{equation}
which completes the construction of $O_f$ in \figref{fig:big-picture}.


\section{Performance estimation}
\label{sec:performance}

In this section we estimate the time and space complexity, and make a comparison between quantum ray tracing and classical path tracing.

First, we discuss the complexity of quantum ray tracing with respect to parameters like scene complexity $P$, maximum depth $D$, number of scattered rays at each intersection $R$, the comparator precision $C$, the precision in quantum counting $T$.

From Eq.~\eqref{eq:id-register} we know the size of $id$ register is $\log R+\log C$.
In the implementation of $O_f$ in \figref{fig:big-picture}, the space and time complexity of a single $FindIntersection$ procedure is $O(P)$, so the space and time complexity of the whole $O_f$ is $O(DP+\log R+\log C)$.
Moreover, there are $\log T$ more qubits required in the quantum counting procedure, thus the overall space complexity is $O(DP+\log R+\log C+\log T)$.

As for the time complexity, since the $G$ in \figref{fig:big-picture} are repeated for $\sum_{j=0}^{t-1}2^j=T-1$ times, and the $QFT^\dagger$ procedure takes $O((\log T)^2)$ time, the overall time complexity is $O(T(DP+\log R+\log C))$.

To compare with classical ray tracing, we should formulate a connection between those parameters above and the precision of results.
If we view the calculation of the color of a single pixel as an integral, more precisely a $2D$-dimensional integral, then our quantum ray tracing algorithm densely samples $R$ points and uses their average to estimate the integral.
The corresponding truncation error is $
    E_R=O(R^{-1/2D})
$, so we choose $
    R=\Omega(E_R^{-2D})
$.
Plus, the quantum counting procedure brings an additional probabilistic error.
From the book\cite{qcqi} we know that if we hope to get a counting estimation with accuracy $E_T$ with success probability at least $1-\epsilon$, then we should choose $
    T = \Omega(1/(E_T\cdot\epsilon))
$.
The truncation error brought by the comparator interval is $E_C=O(1/C)$, thus we choose $
    C = \Omega(1 / E_C)
$.
Therefore, the space complexity becomes,
\begin{equation}
    O\left(DP - 2D\log E_R - \log E_C - \frac{1}{E_T\cdot\epsilon}\right),
\end{equation}
and the time complexity becomes,
\begin{equation}
    O\left(\frac{DP  - 2D\log E_R - \log E_C}{E_T\cdot\epsilon}\right).
\end{equation}

We can see that the major error in $E=E_R+E_T+E_C$ comes from $E_T$, since $E_T^{-1}$ has linear impacts on space and time complexity, while others have logarithm impacts.
For simplicity of analysis we write the space complexity as $
    O(DP-1/(E\cdot\epsilon))
$, and the time complexity as $
    O(DP/(E\cdot\epsilon))
$.

In comparison, the core of path tracing is the Monte Carlo integration.
So if we hope to get an estimation within accuracy $E$ with success probability at least $1-\epsilon$, then the number of rays required is $
    R = \Omega(1/(E^2\cdot\epsilon))
$.
Moreover, the depth of a classical ray path could be dynamic, and modern classical ray tracing algorithm uses various acceleration approaches in the intersection test step to avoid traversing all primitives, like BSP tree\cite{Fuchs1980}, KD-tree\cite{Bentley1975, hapala2011kd} and BVH\cite{Clark1976}.
Assume that the average depth is $D$, and the average number of ray-primitive intersection in searching the nearest intersection of a ray and the whole scene is $O(P)$, then the time complexity is $
    O(RDP) = O(DP/(E^2\cdot\epsilon))
$.
So the quantum ray tracing achieves a quadratic speedup in the sense of being controlled by the same error order.


\section{Conclusion and Discussion}
\label{sec:discussion}

In this paper, we propose a quantum ray tracing algorithm, by first constructing a linear circuit for calculating ray colors, then applying quantum averaging algorithm to extract their mean value.
Finally, we theoretically compare the performances of quantum ray tracing and classical path tracing, and do some simulated experiments to roughly prove our idea.
Unfortunately, it is impossible at present to fully simulate the quantum ray tracing algorithm on classical computers, because it requires exponentially more classical computational resources to simulate quantum computers.
Moreover, the real quantum computers at present cannot provide enough memory to run the algorithm.
We are looking forward to testing our algorithm in a future quantum computer some day.

There are also potential improvements in our work.
In this paper we assume that the time cost for different algorithms to find the intersection of a ray and all scene objects are the same, and compare the time cost of them by counting the number of rays.
Indeed, we overestimate the time cost of modern ray tracing algorithm, since it involves many vital acceleration approaches.
In addition, our quantum ray tracing algorithm uses a fixed-depth ray tree, which may cause visual artifacts and must be compensated.
In classical ray tracing, the problem can be solved by introducing a Russian roulette in the recursion such that there is a probabilistic halting test at each depth.
Such dynamic ray-tree can make the Monte Carlo estimation unbiased.
But the same solution is hard to be implemented in our quantum ray tracing.

In the end, since computer graphics is an application field that requires huge computational power, we hope this paper can be an inspiration of quantum graphics, which studies the quantum solution for more computer graphics problems.

\bibliography{ref}

\begin{thebibliography}{21}%
\makeatletter
\providecommand \@ifxundefined [1]{%
 \@ifx{#1\undefined}
}%
\providecommand \@ifnum [1]{%
 \ifnum #1\expandafter \@firstoftwo
 \else \expandafter \@secondoftwo
 \fi
}%
\providecommand \@ifx [1]{%
 \ifx #1\expandafter \@firstoftwo
 \else \expandafter \@secondoftwo
 \fi
}%
\providecommand \natexlab [1]{#1}%
\providecommand \enquote  [1]{``#1''}%
\providecommand \bibnamefont  [1]{#1}%
\providecommand \bibfnamefont [1]{#1}%
\providecommand \citenamefont [1]{#1}%
\providecommand \href@noop [0]{\@secondoftwo}%
\providecommand \href [0]{\begingroup \@sanitize@url \@href}%
\providecommand \@href[1]{\@@startlink{#1}\@@href}%
\providecommand \@@href[1]{\endgroup#1\@@endlink}%
\providecommand \@sanitize@url [0]{\catcode `\\12\catcode `\$12\catcode
  `\&12\catcode `\#12\catcode `\^12\catcode `\_12\catcode `\%12\relax}%
\providecommand \@@startlink[1]{}%
\providecommand \@@endlink[0]{}%
\providecommand \url  [0]{\begingroup\@sanitize@url \@url }%
\providecommand \@url [1]{\endgroup\@href {#1}{\urlprefix }}%
\providecommand \urlprefix  [0]{URL }%
\providecommand \Eprint [0]{\href }%
\providecommand \doibase [0]{https://doi.org/}%
\providecommand \selectlanguage [0]{\@gobble}%
\providecommand \bibinfo  [0]{\@secondoftwo}%
\providecommand \bibfield  [0]{\@secondoftwo}%
\providecommand \translation [1]{[#1]}%
\providecommand \BibitemOpen [0]{}%
\providecommand \bibitemStop [0]{}%
\providecommand \bibitemNoStop [0]{.\EOS\space}%
\providecommand \EOS [0]{\spacefactor3000\relax}%
\providecommand \BibitemShut  [1]{\csname bibitem#1\endcsname}%
\let\auto@bib@innerbib\@empty
\bibitem [{\citenamefont {Whitted}(1980)}]{Whitted1980}%
  \BibitemOpen
  \bibfield  {author} {\bibinfo {author} {\bibfnamefont {T.}~\bibnamefont
  {Whitted}},\ }\bibfield  {title} {\bibinfo {title} {An improved illumination
  model for shaded display},\ }\href {https://doi.org/10.1145/358876.358882}
  {\bibfield  {journal} {\bibinfo  {journal} {Communications of the ACM}\
  }\textbf {\bibinfo {volume} {23}},\ \bibinfo {pages} {343} (\bibinfo {year}
  {1980})}\BibitemShut {NoStop}%
\bibitem [{\citenamefont {Cook}\ \emph {et~al.}(1984)\citenamefont {Cook},
  \citenamefont {Porter},\ and\ \citenamefont {Carpenter}}]{Cook1984}%
  \BibitemOpen
  \bibfield  {author} {\bibinfo {author} {\bibfnamefont {R.~L.}\ \bibnamefont
  {Cook}}, \bibinfo {author} {\bibfnamefont {T.}~\bibnamefont {Porter}},\ and\
  \bibinfo {author} {\bibfnamefont {L.}~\bibnamefont {Carpenter}},\ }\bibfield
  {title} {\bibinfo {title} {Distributed ray tracing},\ }\href
  {https://doi.org/10.1145/800031.808590} {\bibfield  {journal} {\bibinfo
  {journal} {Computer Graphics}\ }\textbf {\bibinfo {volume} {18}},\ \bibinfo
  {pages} {137} (\bibinfo {year} {1984})}\BibitemShut {NoStop}%
\bibitem [{\citenamefont {Kajiya}(1986)}]{Kajiya1986}%
  \BibitemOpen
  \bibfield  {author} {\bibinfo {author} {\bibfnamefont {J.~T.}\ \bibnamefont
  {Kajiya}},\ }\bibfield  {title} {\bibinfo {title} {The rendering equation},\
  }\href {https://doi.org/10.1145/15922.15902} {\bibfield  {journal} {\bibinfo
  {journal} {Proceedings of the 13th Annual Conference on Computer Graphics and
  Interactive Techniques, SIGGRAPH 1986}\ }\textbf {\bibinfo {volume} {20}},\
  \bibinfo {pages} {143} (\bibinfo {year} {1986})}\BibitemShut {NoStop}%
\bibitem [{\citenamefont {Haines}\ and\ \citenamefont
  {Akenine-M{\"o}ller}(2019)}]{haines2019ray}%
  \BibitemOpen
  \bibfield  {author} {\bibinfo {author} {\bibfnamefont {E.}~\bibnamefont
  {Haines}}\ and\ \bibinfo {author} {\bibfnamefont {T.}~\bibnamefont
  {Akenine-M{\"o}ller}},\ }\href@noop {} {\emph {\bibinfo {title} {Ray Tracing
  Gems: High-Quality and Real-Time Rendering with DXR and Other APIs}}}\
  (\bibinfo  {publisher} {Apress},\ \bibinfo {year} {2019})\BibitemShut
  {NoStop}%
\bibitem [{\citenamefont {Marrs}\ \emph {et~al.}(2021)\citenamefont {Marrs},
  \citenamefont {Shirley},\ and\ \citenamefont {Wald}}]{marrs2021ray}%
  \BibitemOpen
  \bibfield  {author} {\bibinfo {author} {\bibfnamefont {A.}~\bibnamefont
  {Marrs}}, \bibinfo {author} {\bibfnamefont {P.}~\bibnamefont {Shirley}},\
  and\ \bibinfo {author} {\bibfnamefont {I.}~\bibnamefont {Wald}},\ }\href@noop
  {} {\bibinfo {title} {Ray tracing gems ii: Next generation real-time
  rendering with dxr, vulkan, and optix}} (\bibinfo {year} {2021})\BibitemShut
  {NoStop}%
\bibitem [{\citenamefont {Grover}(1997)}]{Grover1997}%
  \BibitemOpen
  \bibfield  {author} {\bibinfo {author} {\bibfnamefont {L.~K.}\ \bibnamefont
  {Grover}},\ }\bibfield  {title} {\bibinfo {title} {Quantum mechanics helps in
  searching for a needle in a haystack},\ }\href
  {https://doi.org/10.1103/PhysRevLett.79.325} {\bibfield  {journal} {\bibinfo
  {journal} {Physical Review Letters}\ }\textbf {\bibinfo {volume} {79}},\
  \bibinfo {pages} {325} (\bibinfo {year} {1997})}\BibitemShut {NoStop}%
\bibitem [{\citenamefont {Shor}(1997)}]{Shor1997}%
  \BibitemOpen
  \bibfield  {author} {\bibinfo {author} {\bibfnamefont {P.~W.}\ \bibnamefont
  {Shor}},\ }\bibfield  {title} {\bibinfo {title} {Polynomial-time algorithms
  for prime factorization and discrete logarithms on a quantum computer},\
  }\href {https://doi.org/10.1137/S0097539795293172} {\bibfield  {journal}
  {\bibinfo  {journal} {SIAM Journal on Computing}\ }\textbf {\bibinfo {volume}
  {26}},\ \bibinfo {pages} {1484} (\bibinfo {year} {1997})}\BibitemShut
  {NoStop}%
\bibitem [{\citenamefont {Lanzagorta}\ and\ \citenamefont
  {Uhlmann}(2005)}]{Lanzagorta2005}%
  \BibitemOpen
  \bibfield  {author} {\bibinfo {author} {\bibfnamefont {M.}~\bibnamefont
  {Lanzagorta}}\ and\ \bibinfo {author} {\bibfnamefont {J.~K.}\ \bibnamefont
  {Uhlmann}},\ }\bibfield  {title} {\bibinfo {title} {Hybrid quantum-classical
  computing with applications to computer graphics},\ }\bibfield  {journal}
  {\bibinfo  {journal} {ACM SIGGRAPH 2005 Courses, SIGGRAPH 2005}\ }\href
  {https://doi.org/10.1145/1198555.1198723} {10.1145/1198555.1198723} (\bibinfo
  {year} {2005})\BibitemShut {NoStop}%
\bibitem [{\citenamefont {Johnston}(2016)}]{Eric2016}%
  \BibitemOpen
  \bibfield  {author} {\bibinfo {author} {\bibfnamefont {E.~R.}\ \bibnamefont
  {Johnston}},\ }\bibfield  {title} {\bibinfo {title} {Quantum supersampling},\
  }\bibfield  {journal} {\bibinfo  {journal} {ACM SIGGRAPH 2016 Talks}\ }\href
  {https://doi.org/10.1145/2897839} {10.1145/2897839} (\bibinfo {year}
  {2016})\BibitemShut {NoStop}%
\bibitem [{\citenamefont {Shimada}\ and\ \citenamefont
  {Hachisuka}(2020)}]{Shimada2020}%
  \BibitemOpen
  \bibfield  {author} {\bibinfo {author} {\bibfnamefont {N.~H.}\ \bibnamefont
  {Shimada}}\ and\ \bibinfo {author} {\bibfnamefont {T.}~\bibnamefont
  {Hachisuka}},\ }\bibfield  {title} {\bibinfo {title} {Quantum coin method for
  numerical integration},\ }\href {https://doi.org/10.1111/cgf.14015}
  {\bibfield  {journal} {\bibinfo  {journal} {Computer Graphics Forum}\
  }\textbf {\bibinfo {volume} {39}},\ \bibinfo {pages} {243} (\bibinfo {year}
  {2020})}\BibitemShut {NoStop}%
\bibitem [{\citenamefont {Abrams}\ and\ \citenamefont
  {Williams}(1999)}]{Abrams1999}%
  \BibitemOpen
  \bibfield  {author} {\bibinfo {author} {\bibfnamefont {D.~S.}\ \bibnamefont
  {Abrams}}\ and\ \bibinfo {author} {\bibfnamefont {C.~P.}\ \bibnamefont
  {Williams}},\ }\bibfield  {title} {\bibinfo {title} {Fast quantum algorithms
  for numerical integrals and stochastic processes},\ }\href
  {https://arxiv.org/abs/quant-ph/9908083v1} {\bibfield  {journal} {\bibinfo
  {journal} {arXiv}\ } (\bibinfo {year} {1999})}\BibitemShut {NoStop}%
\bibitem [{\citenamefont {Alves}\ \emph {et~al.}(2019)\citenamefont {Alves},
  \citenamefont {Santos},\ and\ \citenamefont {Bashford-Rogers}}]{Alves2019}%
  \BibitemOpen
  \bibfield  {author} {\bibinfo {author} {\bibfnamefont {C.~M.}\ \bibnamefont
  {Alves}}, \bibinfo {author} {\bibfnamefont {L.~P.}\ \bibnamefont {Santos}},\
  and\ \bibinfo {author} {\bibfnamefont {T.}~\bibnamefont {Bashford-Rogers}},\
  }\bibfield  {title} {\bibinfo {title} {A quantum algorithm for ray casting
  using an orthographic camera},\ }\href@noop {} {\bibfield  {journal}
  {\bibinfo  {journal} {2019 International Conference on Graphics and
  Interaction (ICGI)}\ ,\ \bibinfo {pages} {56}} (\bibinfo {year}
  {2019})}\BibitemShut {NoStop}%
\bibitem [{\citenamefont {Feynman}(1982)}]{Feynman1982}%
  \BibitemOpen
  \bibfield  {author} {\bibinfo {author} {\bibfnamefont {R.~P.}\ \bibnamefont
  {Feynman}},\ }\bibfield  {title} {\bibinfo {title} {Simulating physics with
  computers},\ }\href {https://doi.org/10.1007/BF02650179} {\bibfield
  {journal} {\bibinfo  {journal} {International Journal of Theoretical Physics
  1982 21:6}\ }\textbf {\bibinfo {volume} {21}},\ \bibinfo {pages} {467}
  (\bibinfo {year} {1982})}\BibitemShut {NoStop}%
\bibitem [{\citenamefont {Feynman}(1986)}]{Feynman1986}%
  \BibitemOpen
  \bibfield  {author} {\bibinfo {author} {\bibfnamefont {R.~P.}\ \bibnamefont
  {Feynman}},\ }\bibfield  {title} {\bibinfo {title} {Quantum mechanical
  computers},\ }\href {https://doi.org/10.1007/BF01886518} {\bibfield
  {journal} {\bibinfo  {journal} {Foundations of Physics 1986 16:6}\ }\textbf
  {\bibinfo {volume} {16}},\ \bibinfo {pages} {507} (\bibinfo {year}
  {1986})}\BibitemShut {NoStop}%
\bibitem [{\citenamefont {Durr}\ and\ \citenamefont {Hoyer}(1996)}]{Durr1996}%
  \BibitemOpen
  \bibfield  {author} {\bibinfo {author} {\bibfnamefont {C.}~\bibnamefont
  {Durr}}\ and\ \bibinfo {author} {\bibfnamefont {P.}~\bibnamefont {Hoyer}},\
  }\bibfield  {title} {\bibinfo {title} {A quantum algorithm for finding the
  minimum},\ }\href {https://arxiv.org/abs/quant-ph/9607014v2} {\bibfield
  {journal} {\bibinfo  {journal} {arXiv}\ } (\bibinfo {year}
  {1996})}\BibitemShut {NoStop}%
\bibitem [{\citenamefont {Brassard}\ \emph {et~al.}(1998)\citenamefont
  {Brassard}, \citenamefont {Hoyer},\ and\ \citenamefont
  {Tapp}}]{QuantumCounting1998}%
  \BibitemOpen
  \bibfield  {author} {\bibinfo {author} {\bibfnamefont {G.}~\bibnamefont
  {Brassard}}, \bibinfo {author} {\bibfnamefont {P.}~\bibnamefont {Hoyer}},\
  and\ \bibinfo {author} {\bibfnamefont {A.}~\bibnamefont {Tapp}},\ }\bibfield
  {title} {\bibinfo {title} {Quantum counting},\ }\href
  {https://doi.org/10.1007/BFb0055105} {\bibfield  {journal} {\bibinfo
  {journal} {Lecture Notes in Computer Science (including subseries Lecture
  Notes in Artificial Intelligence and Lecture Notes in Bioinformatics)}\
  }\textbf {\bibinfo {volume} {1443 LNCS}},\ \bibinfo {pages} {820} (\bibinfo
  {year} {1998})}\BibitemShut {NoStop}%
\bibitem [{\citenamefont {Nielsen}\ and\ \citenamefont {Chuang}(2002)}]{qcqi}%
  \BibitemOpen
  \bibfield  {author} {\bibinfo {author} {\bibfnamefont {M.~A.}\ \bibnamefont
  {Nielsen}}\ and\ \bibinfo {author} {\bibfnamefont {I.}~\bibnamefont
  {Chuang}},\ }\href@noop {} {\bibinfo {title} {Quantum computation and quantum
  information}} (\bibinfo {year} {2002})\BibitemShut {NoStop}%
\bibitem [{\citenamefont {Fuchs}\ \emph {et~al.}(1980)\citenamefont {Fuchs},
  \citenamefont {Kedem},\ and\ \citenamefont {Naylor}}]{Fuchs1980}%
  \BibitemOpen
  \bibfield  {author} {\bibinfo {author} {\bibfnamefont {H.}~\bibnamefont
  {Fuchs}}, \bibinfo {author} {\bibfnamefont {Z.~M.}\ \bibnamefont {Kedem}},\
  and\ \bibinfo {author} {\bibfnamefont {B.~F.}\ \bibnamefont {Naylor}},\
  }\bibfield  {title} {\bibinfo {title} {On visible surface generation by a
  priori tree structures},\ }\href {https://doi.org/10.1145/965105.807481}
  {\bibfield  {journal} {\bibinfo  {journal} {ACM SIGGRAPH Computer Graphics}\
  }\textbf {\bibinfo {volume} {14}},\ \bibinfo {pages} {124} (\bibinfo {year}
  {1980})}\BibitemShut {NoStop}%
\bibitem [{\citenamefont {Bentley}(1975)}]{Bentley1975}%
  \BibitemOpen
  \bibfield  {author} {\bibinfo {author} {\bibfnamefont {J.~L.}\ \bibnamefont
  {Bentley}},\ }\bibfield  {title} {\bibinfo {title} {Multidimensional binary
  search trees used for associative searching},\ }\href
  {https://doi.org/10.1145/361002.361007} {\bibfield  {journal} {\bibinfo
  {journal} {Communications of the ACM}\ }\textbf {\bibinfo {volume} {18}},\
  \bibinfo {pages} {509} (\bibinfo {year} {1975})}\BibitemShut {NoStop}%
\bibitem [{\citenamefont {Hapala}\ and\ \citenamefont
  {Havran}(2011)}]{hapala2011kd}%
  \BibitemOpen
  \bibfield  {author} {\bibinfo {author} {\bibfnamefont {M.}~\bibnamefont
  {Hapala}}\ and\ \bibinfo {author} {\bibfnamefont {V.}~\bibnamefont
  {Havran}},\ }\bibfield  {title} {\bibinfo {title} {Review: Kd-tree traversal
  algorithms for ray tracing},\ }\href
  {https://doi.org/10.1111/j.1467-8659.2010.01844.x} {\bibfield  {journal}
  {\bibinfo  {journal} {Comput. Graph. Forum}\ }\textbf {\bibinfo {volume}
  {30}},\ \bibinfo {pages} {199} (\bibinfo {year} {2011})}\BibitemShut
  {NoStop}%
\bibitem [{\citenamefont {Clark}(1976)}]{Clark1976}%
  \BibitemOpen
  \bibfield  {author} {\bibinfo {author} {\bibfnamefont {J.~H.}\ \bibnamefont
  {Clark}},\ }\bibfield  {title} {\bibinfo {title} {Hierarchical geometric
  models for visible surface algorithms},\ }\href
  {https://doi.org/10.1145/360349.360354} {\bibfield  {journal} {\bibinfo
  {journal} {Communications of the ACM}\ }\textbf {\bibinfo {volume} {19}},\
  \bibinfo {pages} {547} (\bibinfo {year} {1976})}\BibitemShut {NoStop}%
\end{thebibliography}%

\end{document}